\documentstyle[12pt]{article}
\def\centeron#1#2{{\setbox0=\hbox{#1}\setbox1=\hbox{#2}\ifdim
\wd1>\wd0\kern.5\wd1\kern-.5\wd0\fi
\copy0\kern-.5\wd0\kern-.5\wd1\copy1\ifdim\wd0>\wd1
\kern.5\wd0\kern-.5\wd1\fi}}
\def \ltap{\;\centeron{\raise.35ex\hbox{$<$}}{\lower.65ex\hbox{$\sim$}}\;}
\def \gtap{\;\centeron{\raise.35ex\hbox{$>$}}{\lower.65ex\hbox{$\sim$}}\;}
\def \gsim{\mathrel{\gtap}}
\def \lsim{\mathrel{\ltap}}
\begin{document}
\begin{titlepage}

\begin{flushright}
   MSUHEP--940715 \\
   July 15, 1994 \\
\end{flushright}

\begin{center}
  {\large \bf  Multiplicity Distributions and Rapidity Gaps
  }
  \vskip 0.20in
  {\bf Jon Pumplin

  }
\vskip 0.1in
Physics and Astronomy Department \\
Michigan State University\\
East Lansing MI 48824, U.S.A.

\end{center}

\vskip 0.5in

\begin{abstract}

We examine the phenomenology of particle multiplicity
distributions, with special emphasis on the low
multiplicities that are a background to the study of
rapidity gaps.  In particular, we analyze the multiplicity
distribution in a rapidity interval between two jets, using
the {\footnotesize HERWIG} QCD simulation with some
necessary modifications.  The distribution is not of the
``negative binomial'' form, and displays an anomalous
enhancement at zero multiplicity.  Some useful mathematical
tools for working with multiplicity distributions are
presented.  It is  demonstrated that ignoring particles
with $p_\perp < 0.2 \, {\rm GeV/c}$ has theoretical
advantages, in addition to being convenient experimentally.

\end{abstract}

\end{titlepage}

\section {Introduction}
\label{sec:intro}
The probabilities for various numbers of hadrons to be produced
in a high energy collision, in a fixed region of phase space that
is usually defined by a range of pseudo-rapidity, is known as the
{\it multiplicity distribution}.  Attempts have been made to
understand multiplicity distributions on the basis of intuitive
notions of branching and decay of
``clusters'' \cite{vanhove,clusters}.  Approaches with an explicit
basis in QCD have also been made for regimes where an underlying
hard scattering permits perturbative techniques \cite{QCDModels}.
QCD simulation programs such as {\footnotesize HERWIG} \cite{herwig}
include elements of both of these approaches.

Previous work on the multiplicity distribution $\{P_n \}$ has
centered on KNO scaling and its violation \cite{KNO},
``intermittency'' \cite{inter}, and the factorial moments
$\left\langle n \right\rangle$,
$\left\langle n(n-1) \right\rangle$,\dots \cite{factor}.  These
related concepts emphasize average and larger-than-average
multiplicities, which reflect the multiple soft jet production
that is characteristic of QCD at high energy.  In this paper, we
instead focus on the region of low multiplicity.

At the extreme low end of the multiplicity distribution, one
encounters the physics of {\it rapidity gaps}, which can be
defined as regions of length $\Delta y \! \gsim \! 3$ in rapidity
that contain no final-state particles.  Rapidity gaps offer a
unique insight into the workings of QCD.  They can in principle
be made by the exchange of a color-singlet object, such as an
appropriate state of two or more gluons.  They can also be
considered---by definition---to be a facet of the mysterious
{\it pomeron} that also governs elastic and diffractive scattering.

A particularly interesting type of rapidity gap occurs when
the gap lies between two high-$p_\perp$ jets that are widely
separated in rapidity and approximately back-to-back in azimuthal
angle \cite{mueller,bj,delduca}.  In this paper, we study the
multiplicity distribution in a region between two such jets,
resulting from non-pomeron physics; since that is an unavoidable
background to probing rapidity gap physics.

The major detectors CDF and {D\O} at the Tevatron (${\bar p}p$ at
$\sqrt{s} = 1800 \, {\rm GeV}$), and Zeus and H1 at
HERA \cite{HERA} ($e^- p$ at $\sqrt{s} = 300 \, {\rm GeV}$), can
be used to study rapidity gaps experimentally.  However, the range
in pseudo-rapidity where these detectors are most sensitive,
reduced to leave room for jet evidence of hard scattering, is
not very large.\footnote{
  The coverage in $\eta$ could in principle be extended using
  scintillation counters as ``gap detectors''.  A detector
  upgrade of this type should be relatively simple, since
  it is not necessary to have fine segmentation to look for
  zero particles!
}
It is therefore important to estimate the background from
fluctuations in ``normal'' multiparticle production.  This is
the motivation for our study of the multiplicity distribution
at small $n$.

Quantitative results presented in this paper are based on the QCD
Monte Carlo program {\footnotesize HERWIG} \cite{herwig}.  This
program incorporates the color connections between partons, and
therefore includes the natural suppression of rapidity gaps that
is present in QCD, apart from the possibility of coherent
color-singlet exchange.  It therefore  provides a proper model
for the {\it background} to rapidity gap physics.  The simulation
also models the production and decay of many of the known low-mass
hadronic resonances.  These create short-range rapidity correlations
that strongly influence the multiplicity distribution in small
intervals.  The Monte Carlo also provides an opportunity to appraise
the standard practice of substituting the easily-measured
pseudo-rapidity variable
$\eta = \log \cot \frac{\theta}{2}
= \frac{1}{2} \log \left[ (|\vec{p}| + p_z)/(|\vec{p}| - p_z) \right]$
for the more natural true rapidity
$y = \frac{1}{2} \log \left[ (E + p_z)/(E - p_z) \right]$.

There is no guarantee, of course, that Monte Carlo predictions for
the multiplicity are correct.  But it is not unreasonable to expect
that formulae that will be adequate to parametrize the eventual
experiments should be at least flexible enough to fit the simulated
data.  When real data become available, one may hope to tune the
Monte Carlo parameters to improve the accuracy of the simulation.

Parametrizations based on the simulation may also be useful in
correcting actual data for losses due to incomplete acceptance.
This is especially important for the major detectors CDF and
{D\O}, which were not designed to measure particles with transverse
momenta below $\lsim \! 0.2 \, {\rm GeV/c}$.  On the other hand, we
will use the simulation to show in Sect.\ \ref{sec:EdgeEffects}
that it may actually be {\it desirable} to neglect particles with
very low $p_\perp$, since that region is overly sensitive to
contamination by particles produced in the decays of resonances that
are far away in rapidity, and since particles other than photons at
small $|\eta|$ are kinematically suppressed there anyway.

An outline of the paper is as follows.
Sect.\ \ref{sec:tools} introduces some useful mathematical tools
for working with multiplicity distributions---many of which have
been suggested previously \cite{carruthers}.
Sect.\ \ref{sec:montecarlo} describes results from a {\footnotesize
HERWIG} simulation that is loosely applicable to experiments in
progress at CDF and {D\O}.
Sect.\ \ref{sec:EdgeEffects} examines the region of very low
$p_\perp$.
Sect.\ \ref{sec:conclusion} summarizes principal conclusions.

\section {Theoretical Tools}
\label{sec:tools}
\subsection {Generating Function}
\label{sec:gftools}

Our subject is $\{P_n \}$, the set of probabilities to observe
$n$ particles in an event in a selected region of phase space.
The region is generally defined in terms of pseudo-rapidity
$\eta$, or in terms of the Lego variables $\eta$ and azimuthal
angle $\phi$.  The particles are mainly $\pi^\pm$, and $\gamma$
from $\pi^0$ decay, with average transverse momenta of a few
hundred ${\rm MeV/c}$.

The distribution is conveniently represented by the generating
function
\begin{eqnarray}
g(x) \equiv \sum_{n=0}^{\infty} \, P_n \, x^n
\label{eq:genfun}
\end{eqnarray}
which carries all of the information of $\{P_n \} \,$.  The
factorial moments are related to the behavior of $g(x)$
in the limit $x \rightarrow 1$:
\begin{eqnarray}
g(1) &=& 1 \nonumber \\
g^{\prime}(1) &=& \left\langle n \right\rangle  \nonumber \\
g^{\prime \prime}(1) &=& \left\langle n(n-1) \right\rangle
\nonumber \\
g^{\prime \prime \prime}(1) &=& \left\langle n(n-1)(n-2)
\right\rangle  \nonumber \\
&\dots&
\label{eq:gatone}
\end{eqnarray}
Meanwhile, the low multiplicity region we are interested in
is contained in the behavior as $x \rightarrow 0$:
\begin{eqnarray}
g(0) &=& P_0 \nonumber \\
g^{\prime}(0) &=& P_1 \nonumber \\
g^{\prime \prime}(0)/2! &=& P_2  \nonumber \\
g^{\prime \prime \prime}(0)/3! &=& P_3  \nonumber \\
&\dots&
\label{eq:gatzero}
\end{eqnarray}

In principle, $P_n$ is exactly zero beyond some large maximum
$n$, because the energy in the event is finite; so $g(x)$ is a
high-order polynomial.  In practice, however, $P_n$ falls
smoothly and rapidly (perhaps exponentially) at large $n$, and
becomes immeasurably small long before the maximum value is
approached.  Hence it is appropriate to approximate $g(x)$ by
an analytic function, whose infinite series converges at least
out to $|x|=1$ in the complex plane in view of the fact that
$g(1)=1$.

The analytic behavior of $g(x)$ can be useful.  For if one has
an analytic expression for $g$, from a model or simply a
parametrization, a convenient method to calculate the
corresponding probabilities is to integrate $g(x) x^{-n-1}$
numerically around the unit circle in the complex plane and
use Cauchy's theorem to obtain $P_n$.

\subsubsection{Cluster Decay Theorem}
\label{sec:cdtheorem}
The generating function is a convenient tool for analyzing
models in which ``clusters'' decay independently to make the
observed hadrons.  The clusters can be low-mass objects
such as those assumed in QCD Monte Carlo simulations at a
low $Q^2$ non-perturbative scale, or the hypothetical
objects in branching models, or any of the large number
of hadronic resonances that are the immediate ancestors of
most observed hadrons.

The connection is as follows:  if $P_n^{(1)}$ is the
probability to produce $n$ clusters and $P_n^{(2)}$ is the
probability for a cluster to decay into $n$ particles,
then the overall distribution of particles
$\left\{P_n \right\}$ is given by the generating function
relation
\begin{eqnarray}
g(x) = g^{(1)}(g^{(2)}(x)) \; ,
\label{eq:g1ofg2}
\end{eqnarray}
assuming that the clusters decay independently.  Proof of this
relation follows from the obvious expression:
\begin{eqnarray}
P_n = \sum_{j=0}^\infty P_j^{(1)} \,
\sum_{n_1=0}^\infty P_{n_1}^{(2)} \cdots
\sum_{n_j=0}^\infty P_{n_j}^{(2)} \;
\delta_{n_1 + \cdots + n_j, \,n}  \; .
\label{eq:proof1}
\end{eqnarray}
The proof is easily generalized to show
$g(x) = g^{(1)}(g^{(2)}(g^{(3)}(x)))$
for independent decay of independently-decaying clusters, etc.

A simple but useful special case of this theorem applies to
detection efficiencies.  If $Q < 1$ is the detection
probability (``efficiency'')
for a single particle, one can think of the particle as a
`cluster' with $P_1 = Q$ and $P_0 = 1-Q$.  The effect of the
inefficiency can therefore be expressed by
$g\left[ x \right] \rightarrow g\left[ 1 - Q(1-x) \right]$.

\subsubsection{Independent Sources Theorem}
\label{sec:istheorem}
The generating function is also a convenient tool for analyzing
models in which the observed hadrons come from two or more
statistically independent sources.  An important example
that we will use in Sect.\ \ref{sec:BackgroundEvent} occurs
in simulation programs---and perhaps also in nature---where in
addition to particles resulting from a QCD hard scattering and
its associated radiation, there are particles in the final state
known as the ``soft background event'', from soft interactions
between the other partons in the initial composite hadrons.  This
possibility of background particles leads to the notion of a
{\it survival probability} for rapidity gaps \cite{bj,levin}.
Another example that is important for us are particles that
appear far outside the cone of a jet, as a result of sequential
decays of hadronic resonances produced inside the jet cone.

The relevant theorem is as follows: if two independent sources
have probability distributions $\left\{P_n^{(1)} \right\}$ and
$\left\{P_n^{(2)} \right\}$, then both together result in
\begin{eqnarray}
g(x) = g^{(1)}(x) \times g^{(2)}(x)
\label{eq:g1plusg2}
\end{eqnarray}
Proof follows directly from
\begin{eqnarray}
P_n =
\sum_{n_1=0}^\infty P_{n_1}^{(1)} \,
\sum_{n_2=0}^\infty P_{n_2}^{(2)} \;
\delta_{n_1 + n_2, \, n}  \; .
\label{eq:proof2}
\end{eqnarray}
For the lowest multiplicities, Eq.~(\ref{eq:proof2}) takes
the obvious forms
\begin{eqnarray}
P_0 &=&  P_0^{(1)} \, P_0^{(2)} \label{eq:obvious1} \\
P_1 &=&  P_0^{(1)} \, P_1^{(2)} + P_1^{(1)} \, P_0^{(2)}
 \; .
\label{eq:obvious2}
\end{eqnarray}
The theorem can be generalized to
\begin{eqnarray}
\log g(x) = \log g^{(1)}(x) + \cdots + \log g^{(N)}(x)
\label{eq:LogG}
\end{eqnarray}
for combining $N$ independent sources.  {\it Thus on a
logarithmic scale, generating functions from independent
sources are additive.}

\subsection{Density Function}
\label{sec:densfunc}
Intuitively, we want to make a smooth parametrization of the
multiplicity distribution for $n > 0$, and extrapolate it to
$n=0$ to see if there is an anomalous contribution that would
signal rapidity gap physics.  The parametrization is not a
trivial matter, because $P_n$ varies rapidly with
$n$ at small $n$, especially for large rapidity intervals
where $\left\langle n \right\rangle$ is large.

A representation of the probability distribution that I find
to be useful describes it as a continuous superposition of
Poissons:
\begin{eqnarray}
P_n = \int_0^\infty dz \, \rho(z) \, e^{-z}  z^n / n! \; .
\label{eq:densfun}
\end{eqnarray}
The density function $\rho(z)$ is the relative probability
to have a Poisson process of average multiplicity $z$.
Mathematically, $\rho(z)$ is the Laplace transform of the
generating function:
\begin{eqnarray}
g(1-x) = \int_0^\infty dz \, \rho(z) \, e^{-z \, x} \; .
\label{eq:laplace}
\end{eqnarray}
The moments of the continuous distribution $\rho(z)$ are
the factorial moments:
$\int_0^\infty \rho(z) \, dz = 1$,
$\int_0^\infty \rho(z) \, z \, dz = \left\langle n \right\rangle$,
and in general
\begin{eqnarray}
\int_0^\infty \rho(z) \, z^j \, dz =
\left\langle n(n-1) \cdots (n-j+1) \right\rangle \; .
\end{eqnarray}

The independent sources theorem Eq.~(\ref{eq:g1plusg2}) of
Sect.\ \ref{sec:istheorem} can be expressed in terms of
density functions in the form of a convolution integral
\begin{eqnarray}
\rho(z) = \int_0^z \rho^{(1)}(z_1) \, \rho^{(2)}(z-z_1) \, dz_1
\; .
\label{eq:convol}
\end{eqnarray}

The density function would not of course have to be positive
definite; but it turns out to be so for all distributions
discussed in this paper.  Smoothness of $\rho(z)$ is a good
way to express the physical notion that $P_n$ should be a smooth
function of $n$, with the possible exception of structure at or
near $n=0$ from the rapidity gap physics we wish to study.  The
behavior of $P_n$ at small $n$ is governed mainly by $\rho(z)$
at small $z$.  In the extreme, a term $\propto \! \delta(z)$
would contribute to $P_0$ only.

A convenient way to determine $\rho(z)$ from data in an
experiment or simulation is to fit the data to a parametrization
whose transform is known.  We will do this using a sum of terms
of the form $z^{k-1} \, e^{-b \, z}$, which correspond to the
NBD discussed in Sect.\ \ref{sec:nbd}.
{}From the simulation, we will find empirically that terms with
$k > 1$ describe most of the distribution, so
$\rho(z) \rightarrow 0$ like a power as
$z \rightarrow 0$.  To allow for the possibility that
$\rho(0) \neq 0$, one can also include
a term of the form
\begin{eqnarray}
(1 + b z) e^{-b \, z}
\label{eq:rhozero}
\end{eqnarray}
which has $\rho^{\prime}(0) = 0$.

\subsection{Negative Binomial Distribution}
\label{sec:nbd}
The Negative Binomial Distribution (NBD) is a popular
phenomenological form for multiplicity distributions.
It is defined by
\begin{eqnarray}
P_n =
\left(
\begin{array}{c}
n + k - 1 \\
k - 1
\end{array}\right) \,
\left(\frac{k}{k + \bar n} \right)^k \,
\left(\frac{\bar n}{k + \bar n} \right)^n
\label{eq:nb}
\end{eqnarray}
where
\begin{eqnarray}
\left(
\begin{array}{c}
n + k - 1  \\
k - 1
\end{array}\right) \equiv
\frac{k(k+1)\cdots (k+n-1)}{n!} \; .
\end{eqnarray}
It can be conveniently computed with the recurrence relation
\begin{eqnarray}
P_0 &=& (1 + {\bar n}/k)^{-k} \label{eq:recur1} \\
P_{n+1} &=&
\left(\frac{n+k}{n+1}\right) \,
\left(\frac{\bar n}{\bar n + k} \right) \, P_n \; .
\label{eq:recur2}
\end{eqnarray}
Its factorial moments are given by
$\left\langle n \right\rangle = \bar n$ and in general
\begin{eqnarray}
\left\langle n(n-1) \cdots (n-j+1) \right\rangle =
k(k+1)\cdots (k+j-1) \; \left(\frac{\bar n}{k}\right)^j \;.
\end{eqnarray}
Its generating function is
\begin{eqnarray}
g(x) = \left[ 1 + (1-x){\bar n}/k \right]^{-k} \; .
\label{eq:nbdgen}
\end{eqnarray}
Eq.~(\ref{eq:nbdgen}) implies
\begin{eqnarray}
g^{\prime \prime} \, g / (g^{\prime})^2 = 1 + 1/k \; ,
\label{eq:nbtest1}
\end{eqnarray}
which could be used to test whether a distribution is of the
NBD form.  A related test would be to see if $g / g^\prime$
is a linear function of $x$:
$g(x) / g^\prime(x) = 1/{\bar n} + (1-x)/k$.

The density function for the NBD is
\begin{eqnarray}
\rho(z) = \frac{b^k}{\Gamma(k)} \, z^{k-1} \, e^{-b z}
\label{eq:rhonbd}
\end{eqnarray}
where $b = k / {\bar n}$.  It has
a single peak at $z = {\bar n}(1-1/k)$ if $k>1$, or peaks
at $z=0$ for $0<k<1$.  As described in Sect.\ \ref{sec:densfunc},
a convenient way to
determine $\rho(z)$ from experiment or simulation is to fit
$\{P_n \}$ to a superposition of NBD terms and then use
Eq.~(\ref{eq:rhonbd}) to get $\rho(z)$.

In the limit $k \rightarrow \infty$, the NBD reduces to a
Poisson distribution corresponding to uncorrelated production:
$P_n = ({\bar n}^n / n!) \, e^{-{\bar n}}$,
$g(x) = e^{-(1-x){\bar n}}$,
$\rho(z) = \delta(z-{\bar n})$.

The multiplicity distributions we are interested in display a
single maximum with various degrees of broadness, and fall
rapidly at large $n$.  The two free parameters of the NBD
suffice to fit the first two moments
$\left\langle n \right\rangle = \bar n$ and
$\left\langle n^2 \right\rangle =
{\bar n}^2 (1 + 1/{\bar n} + 1/k)$, and hence
the NBD can provide at least a qualitative description of the
probabilities $P_n$ where they are large.  We will see, however,
that a single NBD does not fit our distributions in detail.

Experimentally, the main published data on multiplicity
distributions in very high energy hadron-hadron collisions
are those of UA5 for charged particles in minimum bias
non-diffractive events at
$\sqrt{s} = 900 \, {\rm GeV}$ \cite{UA5}.  The data for rapidity
intervals $\Delta \eta \leq 2$ are well described by NBD
distributions.  For $2 \leq \Delta \eta \leq 5$, the data are
close to NBD, although the NBD fits are not statistically
acceptable.  However, these data come from only a few thousand
events, and therefore have rather large statistical errors where
$P_n$ is small.  They also have large systematic errors at low
multiplicity, where efficiencies are hard to determine.  Hence
the NBD form might not be magic.  NBD distributions have also been
seen in other data, including $e^+ e^- \rightarrow {\rm hadrons}$
\cite{vanhove,nbdmult} and nucleus-nucleus with low
statistics \cite{jain}.

A systematic study of multiplicity distributions in minimum bias
and/or various hard-scattering processes at the Tevatron has yet
to be carried through, although preliminary results from CDF have
been presented \cite{rimondi}.  Some useful information has
been obtained by E735 \cite{E735}.  It would seem that {D\O}
could directly extend the measurements of UA5 to
$\sqrt{s} = 1800 \, {\rm GeV}$,
since their lack of a magnetic field simplifies the tracking
of charged particles at low momentum.

\section {Monte Carlo Simulation}
\label{sec:montecarlo}
\subsection {Hard Scattering}
\label{sec:hardscattering}
The QCD Monte Carlo program {\footnotesize HERWIG} 5.7 \cite{herwig}
was used to simulate $p \bar p$ scattering at the Fermilab
Tevatron energy $\sqrt{s} = 1800 \, {\rm GeV}$, for final states
that contain two relatively high $p_\perp$ jets separated widely
in rapidity.  We will examine the multiplicity distribution in
the interval between these two ``trigger jets''.

Specifically, we require two jets with $p_{\perp}^{(1)}$,
$p_{\perp}^{(2)} > 30 \, {\rm GeV/c}$,
$-3.5 < \eta_1 < -1.5$, $1.5 < \eta_2 < 3.5$, and
$\vert \eta_2 - \eta_1 \vert > 4$.
We require there to be no additional jet with
$p_\perp > 30 \, {\rm GeV/c}$
elsewhere in the event. The jets are defined
by a cone algorithm that I have used previously \cite{mycone},
with a cone radius of $0.7$ in Lego.  This configuration is
interesting for gap physics.  It is also a good one to study from
an experimental standpoint, because the region between the jets,
in which the multiplicity is to be measured, is in the
best region of the detectors.  Indeed {D\O} has already
published data for a rather similar configuration \cite{Brandt},
and further data from both {D\O} \cite{D0InProgress} and
CDF \cite{CDFInProgress} will be forthcoming.

In leading-order QCD, the exchange of transverse momentum
between the partons that produce the trigger jets is accompanied
by an exchange of color.  As a result, one expects lots of
gluon radiation, and hence average multiplicities greater than
those seen in minimum bias events, in the interval between
the jets.  However, if color-singlet exchanges are significant,
{\it e.g.}, in the form of gluon ladders, one can also expect to
observe some events with rapidity gaps \cite{mueller,bj,delduca}.

{\footnotesize HERWIG} includes all possible QCD $2 \rightarrow 2$
tree diagrams for the hard scattering.  Among these diagrams,
gluon exchange dominates over quark exchange because of
the large rapidity separation and the gluon's higher spin.
Also, the scattering partons are mainly $q$ and $\bar q$
because the large sub-energy
\begin{eqnarray}
\hat s = 2 \, p_{\perp}^{(1)} \, p_{\perp}^{(2)}
\left[\cosh(y^{(1)} - y^{(2)}) - \cos(\phi^{(1)} -
\phi^{(2)})\right]
\cong
p_{\perp}^{(1)} \, p_{\perp}^{(2)} \,
e^{|\eta^{(1)} - \eta^{(2)}|}
\label{eq:shat}
\end{eqnarray}
requires them to have large momentum fractions $x_1$ and
$x_2$, which are suppressed more strongly for gluons by
the parton distribution functions.

{\footnotesize HERWIG} is appropriate for this simulation because
it correctly includes the color structure of the QCD hard scattering.
It also includes the production of many of the actual low-mass
hadronic resonances, which have an important influence on the
multiplicity distribution.  A final important feature is that
the program contains no color-singlet exchange, or pomeron
physics in any other form, so it provides a clean model for
the {\it background} to rapidity gaps.

The most recent version 5.7 of {\footnotesize HERWIG} \cite{herwig}
was used, with its default parameter values except for
${\rm {\scriptstyle PTMIN}} = 30 \, {\rm GeV/c}$ to suit our
desired jet $p_\perp$'s, and ${\rm {\scriptstyle PRSOF}} = 0$
which will be discussed in Sect.\ \ref{sec:BackgroundEvent}.
It was necessary to modify the off-the-shelf program to remove
unphysical behavior that otherwise appears for our rare
final state, as follows.  The underlying $2 \rightarrow 2$
cross section in {\footnotesize HERWIG} is evaluated for
on-mass-shell partons.  But the partons are actually off shell
as a result of the
initial state radiation branchings that are a principal
feature of the program.  To enforce a reasonable consistency,
we reject events for which the squared four-momentum of either
initial parton is larger in magnitude than
\begin{eqnarray}
Q^2 \equiv 2 \, {\hat s} \, {\hat t} \, {\hat u} /
{\hat s}^2 \, {\hat t}^2 \, {\hat u}^2 \; ,
\label{eq:qsq}
\end{eqnarray}
a symmetric measure of the hardness of the scattering.\footnote{
  This modification is necessary to avoid unphysical behavior
  in {\footnotesize HERWIG} 5.7 even though a bug corrected in
  that version improved the situation as compared to version 5.6.
}
Events in which either observed jet axis differs by more than
$1.0$ in rapidity from the scattered parton
(${\rm {\scriptstyle IHEP}}=7,8$) responsible for it are also
rejected.  This cut removes only $10 \%$ of the events.  It
guarantees that the trigger jets, which are the
two largest $p_\perp$ jets in the event, come from the underlying
hard scattering, as they should to be consistent with the
approximations on which the simulation is based.

The {\footnotesize HERWIG} program was modified to improve its
efficiency for generating events that satisfy our cuts, with no
further change in content, by replacing its uniformly random
generation of the two final rapidities in the $2 \rightarrow 2$
subprocess with an appropriately peaked one.  This of course
required the event weighting to be handled by the user's program.

Fig.~1 shows the multiplicity distribution, based on
130,000 Monte Carlo events, for particles with
$p_\perp > 0.2 \, {\rm GeV/c}$ in a rapidity interval of
length $2.5$ centered between the two jets in each event.
The center of the interval, $(\eta^{(1)} + \eta^{(2)})/2$,
is distributed with a mean of $0$ and a standard deviation
of $0.36\,$.  The nominal jet cones lie entirely outside the
interval, since the jet axes are at least
$4.0 = 2.5 + 2\times 0.75$ units apart.

The {\it dashed} curve in Fig.~1 shows an attempt to fit the
distribution with a negative binomial form. Although it has
the correct qualitative behavior, {\it the single NBD does
not accurately represent $\left\{P_n \right\}$.}\footnote{
\protect This result is not inconsistent with a previous
 claim \cite{NBDclaim} that Monte Carlo simulations are NBD.
 That claim is based on only 2000 events, implying large
 statistical errors wherever $P_n$ is small; and even with the
 large statistical errors, many of the fits described are
 inadequate in the sense of $\chi^2$.
}
The parameters of the fit shown ($\bar n = 14.21$, $k = 3.78$)
were chosen to match $\left\langle n \right\rangle$ and
$\left\langle n^2 \right\rangle$, but other choices
don't work much better.  The NBD fit is particularly poor in
the region of small $n$ that is our major interest.

The solid curve in Fig.~1 is a good fit to
$\left\{P_n \right\}$ for all $n \neq 0$.  This fit has a
good $\chi^2$, and continues to fit at larger values of
$n$ (not shown), all the way out to $n \sim 60$, beyond which
statistical errors become overwhelming.  The fit consists of
a sum of two NBD terms.  However, the density function
representation $\rho(z)$ in Fig.~2 (solid curve) shows
that these two terms do not describe distinct peaks, but
rather overlap to form a single very smooth distribution.
The distribution is qualitatively similar to, but somewhat
narrower than, the single NBD approximation (dashed curve).

Although the solid curve in Fig.~1 is a smooth fit to the
$n \neq 0$ data, its extrapolation to $n=0$ is $0.0018$, which
underestimates the actual $P_0 = 0.0035$ by a factor of
$1.9 \,$.  {\it This raises a warning flag for rapidity gap
searches, where the signal would correspond to just such an
``extra'' probability for $n=0$!}  A similar but even stronger
effect occurs if all particles are included instead of just
those with $p_\perp > .2\,$:  the actual value
is $P_0 = 0.0014$ while the extrapolation gives $0.0003$.
Similar behavior also occurs if the interval is defined
using true rapidity $\Delta y = 2.5$ in place of
$\Delta \eta = 2.5\,$:
$P_0 = 0.0011$, ${\rm fit} = 0.0002$;
or for a larger interval such as $\Delta \eta = 3.0$:
$P_0 = 0.0011$, ${\rm fit} = 0.0002$.
The effect is
present but somewhat smaller if
only charged particles with $p_\perp > 0.2$ are counted:
$P_0 = 0.0093$, ${\rm fit} = 0.0064$, or if
charged particles with all $p_\perp$
are counted:  $P_0 = 0.0052$, ${\rm fit} = 0.0039$.
The effect remains if all hadron resonances are made stable
instead of being allowed to decay:  in the interval
$\Delta y = 2.5$ we have $P_0 = 0.0033$, ${\rm fit} = 0.0014$,
or in the longer interval
$\Delta y = 3.0$ we have $P_0 = 0.0016$, ${\rm fit} = 0.0005$.

These results indicate that in order to establish a rapidity
gap signal experimentally, the signal will have to be large
compared to the background estimated by extrapolation
from larger $n$, since the extrapolation can underestimate the
non-pomeron contribution.

The dotted curve in Fig.~2 shows the density function for a
parametrization that fits $\left\{P_n \right\}$ in Fig.~1
for all $n$.  The parameterization contains a term of the form
Eq.~(\ref{eq:rhozero}) that allows the extra probability for $P_0$
to appear as structure in $\rho(z)$ at very small $z$ with
$\rho(0) \neq 0$.

\subsection {Background Event}
\label{sec:BackgroundEvent}
The colliding $p$ and $\bar p$ hadrons are extended objects
containing many partons.  Events of the type we are interested
in, where two partons interact to produce jets, will generally
occur only in collisions for which the impact parameter is small.
There are likely to be additional soft interactions between
other constituents, of the same character as those of the typical
``minimum-bias'' interactions that account for the fact that the
inelastic interaction probability is nearly $1.0$ at small
impact parameter.  These additional interactions lead to the
production of particles known as the ``soft background event'',
which raise $\left\langle n \right\rangle$ and decrease $P_0$.

It is reasonable to assume that the soft background particles
are statistically independent from the particles we have
considered so far.  This allows us to compute the final
$\{P_n \}$ by combining fits to its hard and soft components,
using the $g(x) = g^{(1)}(x) \times g^{(2)}(x)$ theorem
(Eqs.~(\ref{eq:g1plusg2})--(\ref{eq:proof2})) or a simple
Monte Carlo.  This is a significant technical help to the
calculation, since $P_0$ is so small that it would otherwise
require extremely many events from the QCD simulation to
determine it accurately.

The {\footnotesize HERWIG} package contains a model for the soft
background event, which was turned off by setting the parameter
${\rm {\scriptstyle PRSOF}} \! = \! 0$ to obtain the hard
scattering results discussed in Sect.\ \ref{sec:hardscattering}
above.  Turning it on in every event via
${\rm {\scriptstyle PRSOF}} \! = \! 1$ leads to soft background
particle distributions that are well described (at $40,000$ event
statistics) by single NBD distributions, except for a sizable
extra contribution to $P_0$.  The {\footnotesize HERWIG} model
for the background event is based on the UA5 data, so it is perhaps
not surprising that it has an NBD form, although this result is not
obvious, since the model actually assumes an NBD form for
{\it clusters} rather than for final particles.  I have checked
that, at any rate, the model predicts charged-particle
multiplicity distributions consistent with those observed by UA5.
The distributions are rather broad  in the sense that the NBD
parameter $k$ is small.  For example, for particles in the region
$\Delta \eta = 2.5$, $p_\perp > 0.2$ corresponding to Figs.~1--2,
the background NBD parameters are $k = 1.8$ and $\bar n = 15.2$,
with an extra contribution of $0.03$ to $P_0$.  The origin and/or
validity of the ``extra'' contribution, which makes $P_0$ larger
than any other single $P_n$, is unclear; so I have tried computing
the final $\{P_n \}$ both with it and without it.

Including the background event, with the extra contribution
to $P_0$ ``on'', changes the multiplicity distribution of Fig.~1
to that shown in Fig.~3a.  An expanded view showing the details
at small $n$ is shown in Fig.~3b.  The mean
$\left\langle n \right\rangle$ has become larger, since it
is equal to the sum of the means from the hard scattering
and the background; while the probabilities at small $n$ have
become much smaller since, {\it e.g.}, to get the extreme
case $n=0$ there must be no particles in the interval from the
hard scattering and also none from the soft background
event (Eq.~(\ref{eq:obvious1})).  The important qualitative
conclusions of Sect.\ \ref{sec:hardscattering} remain true
with the background event included, however:
\begin{itemize}

\item
$\left\{P_n \right\}$ is quite similar to a single NBD form
(dashed curve, based on $\bar n = 28.88$, $k = 4.54$ which fit
$\left\langle n \right\rangle$ and
$\left\langle n^2 \right\rangle$); but the single NBD does not
provide a fully acceptable fit, and is particularly unsuitable
for describing the low multiplicities.

\item
$\left\{P_n \right\}$ can be fit very well for all but the
lowest values of $n$ by a sum of two NBD terms (solid curve).

\item
The accurate two-NBD fit corresponds to a single smooth
peak in the Poisson density function $\rho(z)$ (Fig.~4),
which is slightly narrower than the single NBD approximation.

\item
$P_0$ is larger than the fit at $n=0$ by
roughly a factor of $2$.

\end{itemize}

Fig.~3 is based on 1,000,000 events.  This large number of events
was used to obtain sufficient statistics to show the behavior at
small $n$ clearly, in view of the rather small values of $P_n$
there.  However, the inadequacy of a single NBD fit already sets
in for $\gsim \! 50,000$ events.

The data in Fig.~3b {\it can} be fit by a single NBD term, with
the normalization treated as a free parameter, over the small
region $1 \leq n \leq 12 \,$.  This provides an alternative way
to extrapolate to $n=0$.  It is consistent with the result of
using the two-NBD fit to the entire $n > 0$ distribution.

The entire distribution in Fig.~3, {\it including} $n=0$, can
be fit by a sum of two NBD terms plus a term of the form
Eq.~(\ref{eq:rhozero}) that allows $\rho(0) \neq 0$.  The
size of this term is in fact so small that its effect would
not be visible in Fig.~4.  Philosophically, however, one
does not expect contributions with finite probability density
at average particle number zero, apart from true rapidity gap
processes.

If, instead of counting particles, one counts cells (``towers'')
of size $0.1 \times 0.1$ in $(\eta, \phi)$ space that have
$E_\perp > 0.2$, as is done in calorimeter detectors, there is
essentially no change in the above results, since these cells
are so small that it is rare for more than one particle to
enter a given cell, even at the higher values of $n$.

An attempt was made to estimate the effect on the multiplicity
distribution of the geometric acceptance and detector efficiency
in the current {D\O} experiment \cite{D0InProgress}.  The
assumed geometric
acceptance was $1.0$ (perfect) for $|\eta| < 1.1$, with a
``hole'' corresponding to the edge of the central calorimeter
($0.0$ for  $|\eta|= 1.2 - 1.4$,
$0.5$ for $|\eta|= 1.1 - 1.2$ and $|\eta|= 1.4 - 1.5$), and
a linear fall-off from $0.7$ at $|\eta|= 1.5$ to $0.1$ at
$|\eta|= 3.2$.  The assumed efficiency for photons rises
steeply from $0$ at $p = 0.2 \, {\rm GeV/c}$ to $0.94$ by
$p = 1 \, {\rm GeV/c}$.  The assumed efficiency for charged
particles rises more slowly, reaching
$0.54$ at $p = 1 \, {\rm GeV/c}$ and
$0.80$ at $p = 3 \, {\rm GeV/c}$.
For intervals of length $\Delta \eta = 3.0$ between the jets,
with soft background particles included, the multiplicity
distribution is close to a single NBD with $\bar n = 14.0$ and
$k = 4.4\,$.  Deviations from the single NBD fit begin to
appear clearly when the number of events is $\gsim \! 50,000$.
A sum of two NBD terms fits the distribution even for
$\sim \! 10^6$ events, with very little anomalous contribution
to $P_0 = 0.0014\,$.

\section {Edge Effects and Transverse Momentum}
\label{sec:EdgeEffects}

Figs.~1--4 are based on counting particles with
$p_\perp > 0.2 \, {\rm GeV/c}$.  One reason to require a
minimum $p_\perp$ is to mimic typical experimental
acceptance.  But by using the QCD simulation to study the
types and origins of particles that contribute to the
multiplicity distribution, we will see that the
$p_\perp$ cut also provides some theoretical benefits.

Let us focus in particular on the interval of length
$\Delta \eta = 2.5$ centered at $\eta = 0 \, \pm \, 0.36$
considered in Sect.\ \ref{sec:hardscattering}.  The composition
of particles is
$51\%$ $\gamma$,
$39\%$ $\pi^\pm$,
 $6\%$ $K^\pm, K_L^0$, and
 $3\%$ $p, \bar p, n, \bar n$.
The flood of photons is caused by the fact that {\it true}
rapidity is the proper measure of longitudinal phase space.
The Lorentz frame-specific pseudo-rapidity $\eta$ and the true
rapidity $y$ are related by
\begin{eqnarray}
\sinh \eta &=& p_z / p_\perp \\
\sinh y &=& p_z / \sqrt{p_\perp^2 + m^2} \; .
\label{eq:yveta}
\end{eqnarray}
These become equal for $|p_\perp| \gg m$, as is always the
case for the massless photon; while particles whose transverse
momentum is small compared to their masses are swept out to
$|\eta| > |y|\,$.\footnote{
 \protect Attempts \cite{CDFdNdETA,MIMI} to measure the average
 charged multiplicity $dN_{\rm ch}/d\eta$ apparently have
 ignored this pseudo-rapidity phenomenon when extrapolating
 to account for the unmeasured portion of the spectrum at
 small $p_\perp$.  They also use a phenomenological form
 $dN/d\eta \, dp_\perp^2 \propto (p_\perp + p_0)^{-a}$
 which has incorrect analytic behavior in $p_\perp^2$:
 $dN/dy \, dp_\perp^2 \propto (p_\perp^2 + p_0^2)^{-a}$
 would be preferable.
}
Imposing the cut $p_\perp > 0.2 \, {\rm GeV/c}$ eliminates
much of this pseudo-rapidity effect, and changes the
composition to
$34\%$ $\gamma$,
$52\%$ $\pi^\pm$,
 $9\%$  $K^\pm, K_L^0$, and
 $5\%$  $p, \bar p, n, \bar n$.

A benefit of imposing the minimum $p_\perp$ cut, in addition
to experimental convenience, is that it tends to eliminate
particles whose pseudo-rapidities are unrepresentative
of the actual underlying physics.  For example, low $p_\perp$
photons generally come from the decay of a $\pi^0$ whose true
rapidity differs from that of the photon by something on the
order of $0.5$ units.  Furthermore, the $\pi^0$ may come from
the decay of a $\rho^\pm$ or other low-mass resonance that is
still further away; and that resonance may itself be a resonance
decay product.  From the standpoint of rapidity gap physics, one
would like to think of $\pi$, $\rho$, or higher-mass resonances
alike as stable. Decay effects are simply a source of ``noise''
in the measurement of the rapidity of produced hadrons.  This
noise is significant because the intervals we are looking at
are not very many units long in rapidity.

The {\footnotesize HERWIG} simulation leads to the following
quantitative results for the $\Delta \eta = 2.5$ interval.  If
all particles are included, $16\%$ come from parent particles
or resonances produced outside the corresponding interval
$\Delta y = 2.5$ of true rapidity.  This includes $5\%$ from
parents that are more than $0.5$ units outside the interval.
Imposing the cut $p_\perp > 0.2 \, {\rm GeV/c}$ improves
the situation considerably:  only $9\%$ come from parents
outside the true interval $\Delta y = 2.5$, with only $1\%$
more than $0.5$ units outside it.

Rapidity gaps are traditionally defined by a total absence of
particles in a particular interval of pseudo-rapidity.  We
should not be single-minded about this, however, since as shown
above, neglecting particles with
$p_\perp < p_\perp^{\rm MIN} \approx 0.2 \, {\rm GeV/c}$
substantially improves the connection between the
pseudo-rapidities of the long-lived or stable particles that
are measured and the true rapidities of their parent hadrons.

An additional motive for neglecting low $p_\perp$ particles,
aside from experimental practicality, is the following:  the
theoretically significant variables in rapidity gap physics are
generally the invariant masses of the hadronic systems to the
left and right of the gap.  These masses depend on the ``$+$''
and ``$-$'' components of light-cone momentum,
$\sqrt{p_\perp^2 + m^2} \; e^{\, \pm \, y}$, for which
particles with very low $p_\perp$ are less important.

\section {Conclusion}
\label{sec:conclusion}

The ideal way to look for rapidity gaps would be to measure
the multiplicity zero component $P_0$ as a function of
interval size $\Delta \eta$.  The background from non-gap
physics should decline rapidly with increasing
$\Delta \eta$---for example,
$P_0 \propto e^{-2 \, \Delta \eta}$ roughly
describes the results of our simulation in the range
$\Delta \eta \approx 1 - 3$.  Any residual constant or
slowly varying component at large $\Delta \eta$ is the
rapidity gap signal.

Because current experiments are limited in interval size, and
because it is necessary to make experimental corrections based
on measurements at non-zero $n$, we have studied instead the
form of the $P_n$ distribution in fixed regions of
$\Delta \eta$.  One can hope that for $\Delta \eta \simeq 3$,
the rapidity gap signal will appear as an anomalously large
contribution to $P_0$, when compared to a smooth parameterization
that describes the rest of the distribution.

With the help of the QCD simulation program
{\footnotesize HERWIG}, we have found suitable ways to parametrize
the $P_n$ distribution.  A particularly convenient choice is a sum
of two NBD terms, which gives a very good fit, and automatically
provides a simple parametrization of the generating function $g(x)$
and the Poisson density function $\rho(z)\,$.  The smoothness of the
parametrization is demonstrated by the smooth single-peak form of
$\rho(z)$ (Figs.~2, 4).  The absence of an anomalous contribution
(or a gap signal) at very small $n$ can be characterized by a
power-law behavior $\rho(z) \sim {\rm const} \times z^a$ where
$a>0$ so that $\rho(0) = 0$.  Meanwhile, the often-used single NBD
form has this property, and describes the distribution qualitatively
(Fig.~3a); but is not good enough at small $n$ (Fig.~3b) for
measuring the background to a rapidity gap signal unless the
region of $n$ included in the fit is sharply restricted.

We find from the simulation that fits to $P_n$ for $n \geq 2$
can underestimate $P_0$ by a factor $\sim \! 2$.  This is a
cautionary tale for rapidity gap studies, because the excess
$P_0$ has the same form as a rapidity gap signal.  It will
nevertheless be possible to measure true gap effects in
intervals as small as $2-3$ units, {\it provided} the $n=0$
cross section turns out to be large compared to the extrapolation
from larger $n$.  This will happen if the signal turns out to be
on the order of $1\%$ or larger.  Results from the experiments
are eagerly awaited!

Further work is needed to create a phenomenological model to
describe the rapidity gap physics itself, which contributes
not only to $n=0$, but also to other low multiplicities since
the gap in a given event can be just slightly shorter than
the rapidity interval under consideration, and since there are
edge effects associated with resonance decays as discussed in
Sect.\ \ref{sec:EdgeEffects}.

\section*{Acknowledgments}

I wish to thank J. D. Bjorken for discussions of rapidity
gap physics, and H. Weerts, A. Brandt and S. Kuhlmann for
discussions of the experimental situation.  This work was
supported in part by the Texas National Laboratory Research
Commission grant to the CTEQ collaboration.


\clearpage
\section*{Figure Captions}

\begin{enumerate}

\item
Multiplicity distribution for particles with $p_\perp > 0.2$
in a pseudo-rapidity interval $\Delta \eta = 2.5$ centered
between jets, with ``soft background event'' turned off.
{\it Solid curve:}  good fit for $n \geq 1$;
{\it dashed curve:} single NBD fit.

\item
Poisson density function representations $\rho(z)$ of the
fits shown in Fig.~1.  {\it Dotted curve:}  density function
of fit to all data, including $n = 0$, in Fig.~1.

\item
(a) Multiplicity distribution similar to Fig.~1, with
``soft background event'' included.
(b) Expanded view of the distribution at small $n$.
{\it Solid curve:}  good fit for $n \geq 2$;
{\it dashed curve:} single NBD fit.

\item
Poisson density function representations of the fits in Fig.~3.

\end{enumerate}
\end{document}